\documentclass[aip,pop,preprint,superscriptaddress]{revtex4-1}
\usepackage{amsmath, amssymb}
\usepackage{bm}
\usepackage[pdftex]{graphicx}
\usepackage{color}
\usepackage{hyperref}
\hypersetup{linktoc=page}
\hypersetup{
    colorlinks,
    citecolor=blue,    filecolor=blue,
    linkcolor=blue,     urlcolor=blue
}
\begin{document}
\title{Kinetic plasma waves carrying orbital angular momentum}
\author{D. Blackman}
\author{R. Nuter}
\affiliation{CELIA, University of Bordeaux, CNRS, CEA, 33405 Talence, France}
\author{Ph. Korneev}
\affiliation{National Research Nuclear University ``MEPhI'' (Moscow Engineering Physics Institute), Moscow, 115409 Russia}
\affiliation{P. N. Lebedev Physics Institute, Russian Academy of Sciences, 119991 Moscow, Russia}
\author{V.T. Tikhonchuk}
\affiliation{CELIA, University of Bordeaux, CNRS, CEA, 33405 Talence, France}
\affiliation{ELI-Beamlines, Institute of Physics, Czech Academy of Sciences,  25241 Doln\'{i} B\v{r}e\v{z}any, Czech Republic}

\begin{abstract}
The structure of Langmuir plasma waves carrying a finite angular orbital momentum is revised in the paraxial optics approximation. It is shown that the kinetic effects related to higher-order momenta of the electron distribution function lead to coupling of Laguerre-Gaussian modes and result in modification of the wave dispersion and damping. The theoretical analysis is compared to the three-dimensional particle-in-cell numerical simulations for a mode with orbital momentum $l=2$. It is demonstrated that propagation of such a plasma wave is accompanied with generation of quasi-static axial and azimuthal magnetic fields which are consequence of the longitudinal and orbital momentum transported with the wave. 
\end{abstract} 

\maketitle

\section{Introduction}\label{sec1}
It was discussed in the seminal paper by Allen et al.~\cite{Allen92} that electromagnetic waves may carry orbital angular momentum (OAM), while propagating in vacuum, which can be transferred to particles if the wave is absorbed. This feature has found various applications in optics for compact storing of information, nano-scale imaging and manipulation~\cite{Zhan09}. Mathematically such beams are presented with Laguerre-Gaussian functions, which are eigenmodes of the paraxial optics equation in the cylindrical coordinates. Recent publications show the potential applications of OAM modes in particle focusing and acceleration, generation of strong plasma waves, wake-field excitation, and quasi-static magnetic fields~\cite{Vieira14, Andreev16, Vieira18, Nuter18}

Propagation of OAM optical beams in plasmas is associated with excitation of plasma waves that may also carry orbital momentum~\cite{Mendonca09}. The study of these waves is of particular interest as they are coupled to plasma electrons and are involved in such processes as Landau damping and particle acceleration.  The kinetic plasma waves in the cylindrical geometry have been studied by Mendon\c{c}a~\cite{Mendonca12}. In contrast from common plasma waves these ``twisted plasmons'' demonstrate different dispersion and damping properties. However, the development of the wave dispersion equation in Ref. \cite{Mendonca12} suffers from some inconsistencies and properties of these twisted modes are not sufficiently analyzed. The present paper is dedicated to a more detailed and consistent analysis of the twisted kinetic plasma waves. It is shown that because of direct coupling of plasma wave electric field to particles, the Laguerre-Gaussian (LG) functions are not the eigenfunctions of the electron kinetic equation. While the Poisson equation for the plasma wave electric field can be developed in a series of LG functions, they are coupled in the electron kinetic equation because of the electron motion in the radial and azimuthal direction.  This coupling can be treated by using an expansion on the paraxial parameter - the ratio of the plasma wavelength to the radial width of the wave packet - which is supposed to be small.

An additional area of interest is the generation of quasi-static magnetic field on the second order of the amplitude of the plasmon. This phenomenon was previously observed in simulations described in Ref. \cite{Shi18} where two co-propagating OAM laser pulses with differing angular mode, frequency, and wavelength are injected into a plasma and couple with an OAM plasmon. The resulting plasmon is shown to generate a second-order quasi-static magnetic field. The distribution function obtained in the analysis performed here is used to calculate the second-order magnetic field. The resulting field structure is significantly more complex than the field described in Ref. \cite{Shi18}, we also present numerical results which match the theoretical predictions made here.

In what follows we briefly recall the representation of the paraxial optics wave equation in a series of LG functions and apply the same approach to electron plasma wave equation. It is shown that the modes with different radial and azimuthal wave numbers are coupled to each other, so no definite angular momentum can be associated with a plasma wave. However, in the paraxial approximation, where the plasma wavelength is much smaller than the radius of the wave beam in the transverse direction, only the coupling between neighbouring modes can be retained and the dispersion equation can be presented in a closed form. 

The analysis of this dispersion equation in several particular cases provides the examples of the specific evolution of twisted plasma waves and their coupling to plasma particles. The analytical results are compared and illustrated with intensive numerical simulations.

\section{Dispersion equation for the plasma wave in a cylindrical geometry}\label{sec2}
\subsection{LG modes in optics}\label{sec21}
An electric field $\textbf{E}$ of the electromagnetic wave propagating in vacuum along $z$-axis can be represented in an envelope approximation as
\begin{equation} \label{eq21}
 {\bf E}={\bf e}\, E_0(\tau)\,\exp(-i\omega t+ikz)\, U(z, r,\theta)
 \end{equation}
where ${\bf e}$ is the constant polarization unitary vector, $\omega$ is the wave frequency, $k=\omega/c$ is the axial wave number, $\tau=t -z/c$ is the co-propagating time, $E_0(\tau)$ is the slowly changing in time amplitude, and a scalar function $U$ is describing the waveform in the transverse plane. It is a solution of the paraxial wave equation 
\begin{equation} \label{eq22}
 (2ik\partial_z+\nabla^2_\perp)\,U=0,
 \end{equation}
where the second derivative is neglected assuming that function $U$ evolves slowly in the propagation direction. In the cylindrical geometry the function $U$ can be developed in a series of eigenmodes, which are the LG functions: 
\begin{equation} \label{eq23}
U(z,r,\theta)= \sum_{p,l} c_{p,l}F_{p,l}(X) \,\exp\left(il\theta+i\varphi_{p,l}+\frac{ikr^2}{2f}\right).
 \end{equation}
Here $X=r^2/w^2$ is the normalized radial coordinate, $w(z)=w_0\sqrt{1+z^2/z_R^2}$ is the beam radius, $w_0$ is the beam waist, $z_R=kw_0^2$ is the Rayleigh length, $\varphi_{p,l}(z)=-(2p+|l|+1)\arctan(z/z_R)$ is the Gouy phase, $f(z)=z+z_R^2/z$ is the wavefront curvature, and $c_{p,l}$ is a constant coefficient. The radial wave number $p\geq0$ is an integer that numerates radial modes. The integer $l$ could be positive or negative, and it numerates the orbital angular momentum (OAM). 

The eigenfunction $F_{p,l}$ is the LG mode:
\begin{equation} \label{eq24}
F_{p,l}(X)= \sqrt{\frac{p!}{(|l|+p)!}} X^{|l|/2} L_p^{|l|}(X) \,{\rm e}^{-X/2}.
\end{equation}
where $L_p^{|l|}(X)$ is a generalized, or associated, Laguerre polynomial of degree $p$ and $l$ which may be defined by the Rodriguez representation \cite{Grad}:
\begin{equation} \label{eq45}
L_p^l(x)=(p!)^{-1}{\rm e}^x x^{-l}d_x^p \left({\rm e}^{-x} x^{l+p}  \right).
\end{equation}
The set of functions $F_{p,l}$ are orthogonal and normalized according to the following relation
\begin{equation} \label{eq25}
\int _0^\infty dX \,F_{p,l}(X) \, F_{p',l}(X)= \delta_{p,p'},
 \end{equation}
where $\delta_{p,p'}$ is the symbol of Kronecker. The orthogonality on different angular momenta $l$ and $l'$ is assured by the factor ${\rm e}^{il\theta}$ in Eq.~\eqref{eq23}. The eigenfunctions $F_{p,l}$ do not depend on the sign of the OAM.

So a relatively simple and compact representation of the OAM beam in optics in vacuum, or in a dielectric medium without spatial dispersion, originates from Eq. \eqref{eq22}, the Laplacian in the transverse plane comes from Maxwell's equations.  Presentation of the wave field in a series of LG functions \eqref{eq23} is valid if the paraxial parameter is sufficiently small,  $1/kw_0\ll 1$. Application of the same approach to the electrostatic electron plasma wave is presented in the next section.

\subsection{LG modes presentation for the plasma wave}\label{sec22}
We consider a small amplitude plasma wave in a constant density plasma described by the electrostatic potential $\Phi$ and the electron distribution function $f_e$. The potential satisfies the Poisson equation
\begin{equation} \label{eq26}
\Delta\Phi= \frac{e}{\epsilon_0} \,\delta n_e
 \end{equation} 
where $e$ is the electron charge, $\epsilon_0$ is the vacuum dielectric permittivity and $\delta n_e=\int d{\bf v}\,\delta f_e$ is the perturbation of the electron density. The potential $\Phi$ is related to the deviation of the electron distribution function $\delta f_e=f_e-f_{e0}$ from the equilibrium Maxwellian distribution 
\begin{equation}
f_{e0} (\mathbf r, \mathbf v, t)= n_{e0} (2\pi T_e/m_e)^{-3/2}\exp(-\varepsilon/T_e),
\label{equlibrium_f0}
\end{equation}
which is characterized by the density $n_{e0}$, temperature $T_e$, electron energy $\varepsilon=m_e{\bf v}^2/2$ and the electron mass $m_e$. 

For a monochromatic plasma wave, with the frequency $\omega$ and wave number $k$, we are looking for solutions to the Poisson equation \eqref{eq26} and the linearized Vlasov kinetic equation in the paraxial approximation, $1/kw_{0}\ll 1$. Following the approach proposed by Mendon\c{c}a~\cite{Mendonca12}, we represent the solution of this system as a series of LG functions:
\begin{eqnarray}
  \Phi(z,r,\theta,t)&&= \sum_{p,l} \phi_{p,l}F_{p,l}(X)\exp\left(-i\omega t +ikz+il\theta+i\varphi_{p,l}+iqX\right),    \label{eq29} \\
  \delta f_e(z,r,\theta,{\bf v},t)&&= \sum_{p,l} f_{p,l}({\bf v})F_{p,l}(X)\exp\left(-i\omega t +ikz+il\theta+i\varphi_{p,l}+iqX\right) \label{eq30},
\end{eqnarray}
where $q=kw_b^2/2f=z/2z_R$ is the factor accounting for the front curvature. Similar to the use of the solution set shown by equation \eqref{eq23} to solve equation \eqref{eq22}, using the set of equations \eqref{eq29} to solve the Poisson equation in the paraxial approximation gives:
%% \begin{equation} \label{eq27}
%%   (2ik\partial_z+ \nabla_\perp^2)\,\Phi=k^2\Phi+ \frac{e}{\epsilon_0} \int d{\bf v}\,\delta f_e.  
%%  \end{equation} 
%% Which reduces to:
\begin{equation} \label{eq27}
  k^2\Phi = - \frac{e}{\epsilon_0} \int d{\bf v}\,\delta f_e.  
 \end{equation} 
By substituting expressions \eqref{eq29} and \eqref{eq30} in that equation one transforms it in a system of algebraic equations for the potential amplitudes $\phi_{p,l}$ and partial distribution functions $f_{p,l} ({\bf v})$. The Poisson equation \eqref{eq27} is linear, it thus provides relations between the coefficients of the same mode:  
\begin{equation} \label{eq31}
 \phi_{p,l}= -\frac{e}{\epsilon_0k^2}\int d{\bf v}\, f_{p,l}.
 \end{equation}
The situation is more complicated with the Vlasov equation, which does not separate into a set of independent equations because the gradient operators $v_z\partial_z$ and $ {\bf v}_\perp\cdot\nabla_\perp$ couple the modes with different orbital momenta and radial structure. The axial derivative can be presented as follows:
\begin{equation}
  {\rm e}^{-i\varphi_{p,l}-iqX} v_z\partial_z {\rm e}^{i\varphi_{p,l}+iqX}F(X)=-(2p+|l|)\frac{iv_z}{kw_b^2}\,F+\frac{iv_z}{2kw_b^2}\,XF -\frac{v_z}{kw_b^2}\, XF^\prime,\nonumber
  \end{equation}
where $F'=dF/dX$. All the terms in the right hand side are of the second order over the paraxial parameter $1/kw_b$ with respect to the dominant term $ikv_zF$. These second-order terms are neglected in our analysis. Then the kinetic equation reads:
\begin{equation} \label{eq28}
  -i(\omega-kv_z)\,\delta f_e + {\bf v}_\perp\cdot\nabla_\perp\delta f_e =-iek v_z \Phi\,\partial_{\varepsilon}f_{e0} -e{\bf v}_\perp\cdot\nabla_\perp\Phi\partial_{\varepsilon}f_{e0} ,
 \end{equation} 
where the expression $\partial_{\bf v}f_{e0}=m_e {\bf v}\partial_\varepsilon f_{e0}$ is used for the derivative of the electron distribution function assuming that it depends only on the electron energy. The operator of differentiation on transverse coordinates can be calculated as follows:
\begin{eqnarray}
  &&{\rm e}^{-iqX} {\bf v}_\perp\cdot\nabla_\perp {\rm e}^{il\theta+iqX}F(X)= {\rm e}^{-iqX}\bigg[v_\perp\cos(\theta-\theta_v)\,\partial_r- \frac{v_\perp}{r}\sin(\theta-\theta_v)\,\partial_\theta\bigg] {\rm e}^{il\theta+iqX}F(X)= \nonumber
\end{eqnarray}
\begin{eqnarray}
  &&=\frac{v_\perp}{w_b}{\rm e}^{i(l+1)\theta-i\theta_v}\sqrt{X}F'+\frac{v_\perp}{w_b}{\rm e}^{i(l-1)\theta+i\theta_v}\sqrt{X}F' -\nonumber\\
  &&+iq\frac{v_\perp}{w_b}{\rm e}^{i(l+1)\theta-i\theta_v}\sqrt{X}F+iq \frac{v_\perp}{w_b}{\rm e}^{i(l-1)\theta+i\theta_v}\sqrt{X}F -\nonumber\\
  &&\frac{l}{2\sqrt{X}}\frac{v_\perp}{w_b}{\rm e}^{i(l+1)\theta-i\theta_v}F+\frac{l}{2\sqrt{X}}\frac{v_\perp}{w_b}{\rm e}^{i(l-1)\theta+i\theta_v}F.\nonumber
\end{eqnarray}
  Here $v_r=v_\perp\cos(\theta-\theta_v)$ and $v_\theta=-v_\perp\sin(\theta-\theta_v)$ are the radial and azimuthal components of electron velocity and $\theta_v$ is the angle of the electron velocity in the transverse plane.  By using the proprieties of the Laguerre functions \cite{Grad}, the derivative of the function $F$ can be expressed as: 
\begin{equation} \label{eq34a}
\sqrt{X}\,F'_{p,l}(X)= \frac12 \sqrt{p+1} F_{p+1,l-1}(X) -\frac12 \sqrt{p} F_{p-1,l+1}(X) .
 \end{equation}
See Appendix \ref{appA} for details. By multiplying Eq. \eqref{eq28} by the factor $F_{p',l'}\,\exp(-il'\theta -iqX)$ and performing integration over the transverse coordinates one obtains the following system of algebraic equations for the coefficients $f_{p,l}$:
\begin{equation}  \label{eq35}
 (\omega-kv_z)\,f_{p,l}+i \sum_{p',l'} M_{p,l;p',l'}f_{p',l'} =ek v_z \phi_{p,l} \partial_{\varepsilon}f_{e0}-i e\sum_{p',l'} M_{p,l;p',l'}\phi_{p',l'}\partial_{\varepsilon}f_{e0},
 \end{equation}
where the matrix elements $M_{p,l;p',l'}$ are defined as follows:
\begin{equation} \label{eq32}
M_{p,l;p',l'}=\frac{1}{\pi w_b^2}\int_0^{2\pi} d\theta\int_0^\infty dr\,r \,F_{p,l}(X) {\rm e}^{-il\theta-iqX}\, {\bf v}_\perp\cdot\nabla_\perp{\rm e}^{il'\theta+iqX} F_{p',l'}(X),
 \end{equation}
Performing the integrations in Eq. \eqref{eq32} one finds the following expression for the matrix elements: 
\begin{equation} \label{eq33}
M_{p,l;p',l'}=\frac{v_\perp}{w_b}\left[{\rm e}^{-i\theta_v}\delta_{l,l'+1} K^-_{p,l;p',l'} +{\rm e}^{i\theta_v}\delta_{l,l'-1} K^+_{p,l;p',l'}\right]. 
 \end{equation}
The matrices $K^+$ and $K^-$ describe coupling of the modes with neighbouring orbital moments:
\begin{eqnarray}
K^\mp_{p,l;p',l'}&=&\frac{\exp[i(\varphi_{p',l'}-\varphi_{p,l})]}{2} \int_0^\infty dX  \,F_{p,l}(X) \left[\sqrt{p'+1}\,F_{p'+1,l'-1}(X) -\sqrt{p'}\,F_{p'-1,l'+1}(X) \right. \nonumber \\
  && \left. +\frac{iz}{z_R}\sqrt{X}\,F_{p',l'}(X)\mp\frac{l'}{\sqrt{X}}F_{p',l'}(X)\right]. \label{eq35k}
\end{eqnarray}
Considering Eq. \eqref{eq35} one can see the principal difference from the paper by Mendon\c{c}a~\cite{Mendonca12}, where couplings between the neighbouring orbital modes were neglected and the operator  $ {\bf v}_\perp\cdot\nabla_\perp$ was replaced by its average value for each mode separately. This set of equations can be further simplified by developing the elements of the electron distribution function in Fourier series of the velocity angle:
$$ f_{p,l}(\theta_v)= \sum_m  f_{p,l}^{(m)}{\rm e}^{-im\theta_v}.$$
Then by integrating Eq. \eqref{eq35} over the azimuthal velocity angle $\theta_v$ one obtains a series of equations for the moments of the partial distribution function $ f_{p,l}^{(m)}$:
\begin{eqnarray}
  && (\omega-kv_z)\,f_{p,l}^{(m)}+ i \frac{v_\perp}{w_b} \sum_{p'}\bigg[K^-_{p,l;p',l-1}f_{p',l-1}^{(m-1)}+ K^+_{p,l;p',l+1}f_{p',l+1}^{(m+1)} \bigg] =\nonumber \\
  &&=ek v_z \phi_{p,l}\delta_{m,0} \partial_{\varepsilon}f_{e0}-i e \frac{v_\perp}{w_b}\sum_{p'} \bigg[K^-_{p,l;p',l-1}\phi_{p',l-1}\delta_{m,1}+K^+_{p,l;p',l+1}\phi_{p',l+1}\delta_{m,-1}\bigg]\partial_{\varepsilon}f_{e0}.\label{eq36}
\end{eqnarray}
Along with Eq. \eqref{eq31}, which includes the function $f_{p,l}^{(0)}$, this system fully defines linear plasma waves with arbitrary orbital momentum. The LG modes are coupled both in orbital momentum $l$ to close neighbours and in radial number $p$ in the first order on the paraxial parameter. 

\subsection{Dispersion equation for the twisted plasma wave}\label{sec23}
The system of equations \eqref{eq31} and \eqref{eq36}, are obtained in the paraxial approximation ($1/kw_b\ll 1$), could be further simplified. The paraxial approximation implies smallness of the mode-coupling terms. Thus, the equation for $f_{p,l}^{(0)}$ in \eqref{eq36} can be simplified by accounting for coupling to $f_{p,l}^{(\pm1)}$ but neglecting the higher-order harmonics. Then, the equations for the first harmonics read:
\begin{equation} \label{eq38a}
 (\omega-kv_z)\,f_{p,l}^{(\pm1)}=-i \frac{v_\perp}{w_b} \sum_{p'} K^\mp_{p,l;p',l\mp1} \left(f_{p',l\mp1}^{(0)} + e \phi_{p',l\mp1}\partial_{\varepsilon}f_{e0}\right).
\end{equation}
Substituting this expression into Eq. \eqref{eq36} for the harmonic  $f_{p,l}^{(0)}$ one finds:
\begin{equation} \label{eq38b}
 f_{p,l}^{(0)} = e \frac{k v_z}{\omega-kv_z} \phi_{p,l} \partial_{\varepsilon}f_{e0}- \frac{v_\perp^2}{w_b^2}\frac{1}{(\omega-kv_z)^2}\sum_{p'}  Q_{p,p'}^{(l)} \left(f_{p',l}^{(0)}+e\phi_{p',l} \partial_{\varepsilon}f_{e0}\right) ,
\end{equation} 
where the notation for the mode-coupling coefficient is introduced:
\begin{equation}\label{eq38}
    Q_{p,p'}^{(l)} =  \sum_{p''\geq0}\bigg[K^-_{p,l;p'',l-1}K^+_{p'',l-1;p',l}+ K^+_{p,l;p'',l+1}K^-_{p'',l+1;p',l}\bigg].
 \end{equation}
The second term in the right hand side of Eq. \eqref{eq38b} contains the dominant term with $p'=p$ and all other terms with $p'\neq p$ are of the second order. By retaining the first-order terms one obtains the final expression  for $f_{p,l}^{(0)}$:
\begin{equation} \label{eq39a}
 f_{p,l}^{(0)} =\left[-1+ \frac{\omega \,(\omega-kv_z)}{(\omega-kv_z)^2+Q_{p,p}^{(l)} v_\perp^2/w_b^2} \right]\, e \phi_{p,l} \partial_{\varepsilon}f_{e0}.
 \end{equation}
In the second term, it is important to account for the second order term in the denominator, which shifts the resonance condition $\omega=kv_z$ due to the transverse structure of the plasma wave. By substituting this expression for the electron distribution function in the Poisson equation \eqref{eq31}  the dispersion equation for the twisted plasma wave is obtained:
\begin{equation} \label{eq39}
\epsilon(\omega,k)=1 + \frac{e^2}{\epsilon_0 k^2} \int d{\bf v}\,\left[-1+ \frac{\omega \,(\omega-kv_z)}{(\omega-kv_z)^2+Q_{p,p}^{(l)} v_\perp^2/w_b^2}\right]  \partial_{\varepsilon}f_{e0}=0.
 \end{equation}
The solution of this equation in the limit $\omega\gg k v_{th}$, where $v_{th}$ is the electron thermal velocity, can be found by using a standard expansion procedure. Here we consider the equilibrium distribution function \eqref{equlibrium_f0}, but the expression \eqref{eq39} is more general. A non-equilibrium distribution function may result from the corresponding plasma wave modes. The real part of the dispersion equation \eqref{eq39} then reads:
$$ {\rm Re}[\epsilon(\omega,k)]=1-\frac{\omega_{pe}^2}{\omega^2}\left(1+\frac{3k^2v_{th}^2}{\omega^2}-\frac{2Q_{p,p}^{(l)}}{k^2w_b^2}\right)$$
where $\omega_{pe}=\sqrt{e^2n_{e0}/m_e\epsilon_0}$ is the plasma frequency. The mode-coupling term contributes then to the plasma wave dispersion:
\begin{equation} \label{eq40}
 \omega^2=\omega_{pe}^2\left(1+3k^2\lambda_{De}^2-2Q_{p,p}^{(l)}/k^2w_b^2 \right).
  \end{equation}
Here $\lambda_{De}=v_{te}/\omega_{pe}$ is the Debye length. The last term in the parenthesis could be comparable with the thermal dispersion. As it is shown below in Eq. \eqref{eq46} the coefficients $Q_{p,p}^{(l)}$ are negative and consequently the OAM and final radial extension of the plasma wave increase its dispersion.

By taking the residue in the resonance terms in the right hand side of Eq. \eqref{eq39} one finds an expression for the plasma wave damping. The Landau resonance in the case of plane wave $v_z= \omega/k$ splits into two resonances $v_z^{\pm}= \omega/k \pm  (v_\perp/kw_b)\sqrt{-Q_{p,p}^{(l)}}$ shifted with respect to the axial phase velocity. By taking the residues of these two resonances one finds expression for the imaginary part of the dielectric permittivity:
\begin{equation} \label{eq41}
{\rm Im}[\epsilon(\omega,k)] =\sqrt{\frac\pi2}\frac{\omega_{pe}^2\omega}{k^3\lambda_{De}^3}\,\exp \left(-\frac{\omega^2}{2k^2v_{th}^2}\right)\, R\left( \frac{2\omega}{k^2v_{th}w_b}\sqrt{-Q_{p,p}^{(l)}} \right).
  \end{equation}
Here, the function $R(\xi)=\int_0^\infty du\,u\,\exp(-u^2/2)\,\cosh(u\xi)$ accounts for the OAM contribution.
The corrections due to the orbital momentum of the plasma wave are of the same order to the dispersion and to the damping. The quantitative contribution is defined by the value of the coupling coefficient $Q_{p,p}^{(l)}$. 

Calculation of the coefficients $K^\pm$ is presented in Appendix \ref{appA}. There are only four non-zero terms in the coefficients $K^-$:
\begin{eqnarray} \label{eq46a}
 K^-_{p,l;p,l-1}=-\frac12\left(1-i\frac{z}{z_R}\right)\sqrt{l+p}, \qquad  K^-_{p,l;p+1,l-1}=-\frac12\left(1+i\frac{z}{z_R}\right)\sqrt{p+1}, \nonumber \\
  K^-_{p-1,l+1;p,l}=-\frac12\left(1+i\frac{z}{z_R}\right)\sqrt{p}, \qquad  K^-_{p,l+1;p,l}=-\frac12\left(1-i\frac{z}{z_R}\right)\sqrt{l+p+1}.
 \end{eqnarray}
The corresponding matching coefficients in the series $K^+$ read:
\begin{eqnarray} \label{eq46b}
 K^+_{p,l-1;p,l}=\frac12\left(1+i\frac{z}{z_R}\right)\sqrt{l+p}, \qquad  K^+_{p+1,l-1;p,l}=\frac12\left(1-i\frac{z}{z_R}\right)\sqrt{p+1}, \nonumber \\
  K^+_{p,l;p-1,l+1}=\frac12\left(1-i\frac{z}{z_R}\right)\sqrt{p}, \qquad  K^+_{p,l;p,l+1}=\frac12\left(1+i\frac{z}{z_R}\right)\sqrt{l+p+1}.
 \end{eqnarray}
Summing these coefficients according to Eq. \eqref{eq38} one finds the final expression for the coupling coefficient:
\begin{equation} \label{eq46}
Q_{p,p'}^{(l)} = -\left(1+\frac{z^2}{z_R^2}\right)\,\left(p+\frac{|l|+1}2\right).
  \end{equation}
As one can see, the mode $p,l$ is coupled in general to four neighbouring modes: $p,\,l\pm1$ and $p\pm1,\,l\mp1$. In the case $p=0$ only three modes are coupled: $0,\,l\pm1$ and $1,\,l-1$. Finally, the principal mode $0,0$ is coupled to two modes $0,1$ and $1,-1$. 

All coupling coefficients are negative. This implies, in agreement with qualitative expectations, that presence of OAM increases the plasma wave dispersion and damping. The final expressions can be written as follows:
\begin{eqnarray}
&&\omega^2=\omega_{pe}^2\left(1+3k^2\lambda_{De}^2+ \frac{2p+ |l|+1}{k^2w_{0}^2} \right),  \label{eq44a}  \\
   &&\frac{{\rm Im} \,\omega}{\omega}=-\sqrt{\frac\pi8}\frac1{k^3\lambda_{De}^3}\,\exp \left(-\frac{\omega^2}{2k^2v_{th}^2}\right)\, R\left(2\frac{\sqrt{p+(|l|+1)/2}}{k^2\lambda_{De} w_{0}} \right).\label{eq44b}
 \end{eqnarray}
 Note  that these expressions are rather different from the expressions (26) and (30) for the plasma wave dispersion and damping proposed in Ref.~\cite{Mendonca12}. In the limit of a very wide (almost planar) wave one finds the standard expressions for the dispersion and damping of a plane Langmuir wave. In contrast, in the case of sufficiently narrow beams, where $w_{0}<1/k^2\lambda_{De}$, the OAM corrections dominate, the function $R$ in that limit $\xi\gg 1$ behaves as $R(\xi)\sim\sqrt{2\pi}\,\xi\,\exp(\xi^2/2)$.
 
 \section{Structure of a vortical plasma wave}\label{sec3}
 \subsection{Numerical calculations}
 In order to test the analytical results presented in this article we carry out numerical calculations in a kinetic framework. These numerical calculations are performed using the particle-in-cell (PIC) code OCEAN \cite{NUTER2016664}. A 3D box with dimensions $1200\times1200\times160$ cubic cells with sides of length $(\omega_0/c)\delta x = (\omega_0/c)\lambda_{De} = 0.125\pi$, this is filled with a uniform hydrogen plasma to a density of $n_e/n_c=0.01$ and temperature of $T_e/m_e c^2=1.54\times10^{-3}$, whilst ions are fixed. The boundary condition along the propagation axis is periodic, whilst  in the transverse directions they are absorbing for both fields and particles. In order to facilitate a simple periodic plasma wave with OAM the Gouy phase and front curvature are ignored for this analysis. 
 
To explore the generation of magnetic fields, simulations are run with a perturbation with a mode of $p=0,\,l=2$, and also with a phase velocity $\omega/k=c$ to avoid damping and trapped particles. The length of the box was chosen so that it fits exactly one wavelength with $k=0.1$. The width $w=50$ $c/\omega_0$ being chosen to avoid additional dispersion from the vortex terms in Eq. \eqref{eq28}. 

In order to properly resolve the second-order magnetic fields at non-relativistic amplitudes ($a_0\sim0.2$) more than 100 particles per cell are required to ensure that any generated magnetic field has a larger amplitude than the numerical noise level. The 1D line-out plots shown in the article are filtered using a window function of the form $W(x)=\sin^2(\pi x/N dx)$, with $N=20$ being the number of cells to average. This has the effect of removing high-frequency noise from the results. The results without the filter are also shown for comparison. The PIC simulation shows a plasmon which is stable over more than 20 periods and an initial set-up phase of 10 periods. More details of the set-up and stability of the simulation are described in Appendix \ref{appB}.
 
\begin{figure}[ht]
\includegraphics{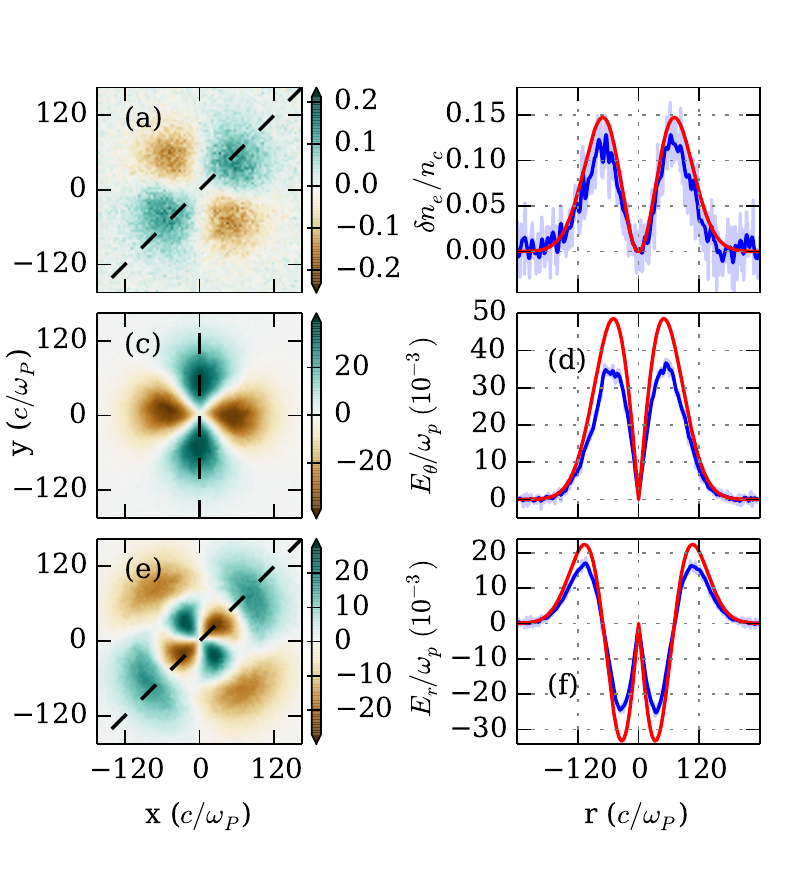}\caption{\label{figpvt} Results from a particle-in-cell simulation 16 periods after the initial 10 period set-up phase. The plots on the top are of $\delta n_e/n_c$, the middle of $E_{\theta}$, and the bottom of $E_r$. The plots on the left (a, c, e) show transverse slices (with no image filter applied) taken from the centre of the PIC code box, with the propagation ($z$) axis going into the page. The dashed lines shown in the transverse slices are the line-outs used to plot the graphics on the right. The plots on the right (b, d, f) are line-outs from the slices (filtered blue, unfiltered light blue) compared with theoretical predictions with an $a_0 = 0.2$ (red).}
\end{figure}
 
\subsection{Electric field of a plasma wave carrying an orbital momentum}\label{sec31}
As an example of LG plasma wave considered in the previous section, we consider here a structure of a single mode $p$, $l$ within the Rayleigh zone $|z|\ll z_R$. The electric potential \eqref{eq29} contains only one term characterized by the amplitude  $\phi_{p,l}$:
\begin{equation} \label{eq51}
 \Phi(z,r,\theta)=  \phi_{p,l}F_{p,l}(X) \,\cos(kz-\omega t +l\theta),     
 \end{equation}
where the radial part is given by the function $F_{p,l}(r^2/w_b^2)$ \eqref{eq24}.  The electric field is found by taking the gradient of the potential:
\begin{eqnarray}
&&   E_z= E_{0} F_{p,l}(X)\, \sin(kz-\omega t +l\theta) , \label{eq52}  \\
&&   E_\theta=\frac{l E_{0}}{kw_b} X^{-1/2} F_{p,l}(X) \, \sin(kz-\omega t +l\theta) , \label{eq53}  \\
&&   E_r=-2 \frac{E_{0}}{kw_b} X^{1/2} F_{p,l}^\prime (X) \, \cos(kz-\omega t +l\theta),  \label{eq54} 
 \end{eqnarray}
where $E_{0}=k\phi_{p,l}$ is the amplitude of the axial electric field. The axial field dominates, the transverse fields are smaller by a factor $1/kw_b\ll1$. The radial field is phase shifted with respect to the azimuthal and axial fields.

It is important to assure that the radial and azimuthal electric fields are not singular at the beam axis. As the radial function behaves at the origin $X\ll1$ as $F_{p,l}\propto X^{|l|/2}$, the fields in question behave as $E_\theta \propto r^{|l|-1}$ and $E_r \propto r^{|l|-1}$. Therefore, for $l=\pm1$ these fields are non-zero at the axis, which presents a problem in a cylindrical system. 
%It is important to know the values of the radial and azimuthal electric fields at the beam axis. As the radial function behaves at the origin $X\ll1$ as $F_{p,l}\propto X^{|l|/2}$, the fields in question behave as $E_\theta \propto r^{|l|-1}$ and $E_r \propto r^{|l|-1}$. Therefore, for $|l|>1$ these fields are zero at the axis. 
Noting that an $l=1$ mode does exist in the work presented in Ref. \cite{Vieira18}, the case $l=\pm 1$ looks particular. Here, both fields are taking final, but indefinite values at $r=0$:
$$ E_{\theta(r=0)}=\pm\frac{E_{0}}{kw_{0}} \, \sqrt{p+1}\,\sin(kz-\omega t \pm\theta),\qquad 
E_{r(r=0)}=-\frac{E_{0}}{kw_{b,0}} \, \sqrt{p+1}\, \cos(kz-\omega t \pm\theta). $$
However, these fields in the Cartesian coordinates are regular:
$$  E_{x(x=0,y=0)}=-\frac{E_{0}}{kw_{0}} \, \sqrt{p+1}\,\cos(kz-\omega t),\qquad 
E_{y(x=0,y=0)}=\pm \frac{E_{0}}{kw_{0}} \, \sqrt{p+1}\, \sin(kz-\omega t).$$
They correspond to the field of a dipole rotating in the clockwise direction for $l=1$.

The later time electron density and electric fields obtained from the PIC simulation (shown in Fig. \ref{figpvt}) match the theoretical values closely, though with slightly smaller amplitudes due to the implementation of the initial conditions in the PIC code. 

\subsection{Electron distribution function in the field of a plasma wave}\label{sec32}
The dominant term in the expansion of the electron distribution function \eqref{eq30} is given by Eq. \eqref{eq39a}. In the first-order expansion over the paraxial parameter $1/kw_b\ll1$, the expression is straightforward:
\begin{equation} \label{eq55}
 f_{p,l}^{(0)} = \frac{k v_z}{\omega-kv_z} e\phi_{p,l} \partial_{\varepsilon}f_{e0}.
 \end{equation}
However, there are other coefficients that are of the first order. This follows from Eq. \eqref{eq38a} by taking into account the fact that non-zero coefficients are given by Eqs. \eqref{eq46a} and \eqref{eq46b}. The three components of the electron distribution function in the first order are the following:
\begin{eqnarray}
f_{p,l\pm1}^{(\pm1)}&&=  \pm\frac{iv_\perp}{2w}\frac{\omega}{(\omega-kv_z)^2} e\phi_{0,1} \partial_{\varepsilon}f_{e0}\sqrt{l+p+\frac{1\pm1}{2}} ,\label{eq56}  \\
f_{p\mp1,l\pm1}^{(\pm1)}&&=-\frac{iv_\perp}{2w}\frac{\omega}{(\omega-kv_z)^2}  e\phi_{0,1} \partial_{\varepsilon}f_{e0}\sqrt{p+\frac{1\mp1}{2}} .\label{eq57} 
 \end{eqnarray}  
With these expressions one can calculate the explicit form of the electron distribution function:
\begin{eqnarray}\label{eq57a} 
  & \delta f_e=&\frac{eE_{0}}{\omega-kv_z}\,\partial_\varepsilon f_{e0}\,\bigg[v_z F_{p,l}(X)\cos(kz-\omega t+l\theta) +\nonumber \\
    &&+ \frac{v_\theta l}{kw_b}\frac{\omega}{\omega-kv_z} X^{-1/2} F_{p,l}(X) \cos(kz-\omega t+l\theta) + \nonumber \\
    && -2\frac{v_r}{kw_b}\frac{\omega}{\omega-kv_z} X^{1/2} F_{p,l}^\prime (X)\sin(kz-\omega t+l\theta)  \bigg].
 \end{eqnarray}
Here the radial and azimuthal electron velocities $v_r=v_\perp \cos(\theta-\theta_v)$ and $v_\theta=-v_\perp \sin(\theta-\theta_v)$ are introduced the same way as in Eqs. \eqref{eq32} and \eqref{eq33}. Coupling of the dominant mode $p=0,\,l=1$ to neighbouring modes results in the appearance of azimuthal and radial electron velocities in the expression for the electron distribution function.

This expression can be used for the calculation of the moments of electron distribution function. The lowest moments, the perturbation of density and electric current, can also be found directly from the Poisson and Ampere equations. Explicit solutions for the electron distribution function for mode $p=0,\,l=2$ are derived in Appendix \ref{appC}. According to Eq. \eqref{eq26}, the density perturbation reads
\begin{equation} \label{eq58} 
\frac{\delta n_e}{n_{e0}}= -\frac{ke E_{0}}{m_e\omega_{pe}^2}F_{p,l}(X) \, \cos(kz-\omega t +l\theta) .
 \end{equation}
The electric current follows from the Ampere relation, ${\bf j}_e=-\epsilon_0 \partial_t\bf E$:
\begin{eqnarray}
&&   j_z=\epsilon_0\omega E_{0}F_{p,l}(X)\, \cos(kz-\omega t +l\theta) , \label{eq59}  \\
&&   j_\theta= \epsilon_0\omega \frac{ l E_{0}}{kw_b} X^{-1/2} F_{p,l}(X) \, \cos(kz-\omega t +l\theta) , \label{eq60}  \\
&&   j_r= 2\epsilon_0 \omega \frac{E_{0}}{kw_b} X^{1/2} F_{p,l}^\prime (X) \, \sin(kz-\omega t +l\theta).  \label{eq61} 
 \end{eqnarray}
The same expressions can be found by integrating the expression \eqref{eq57a} for the electron distribution function and accounting for the dispersion relation $\omega\approx \omega_{pe}$. One can also calculate the orbital momentum carried by electrons in the plasma wave. In the first order on the wave amplitude one finds:
\begin{equation}\label{eqlz}
l_z=m_e r \int d{\bf v}\,v_\theta \delta f_e=-\frac{n_{e0}}{\omega}eE_{0}X^{-1/2} F_{p,l}(X) \, \cos(kz-\omega t +l\theta).
\end{equation}
It oscillates in space and time and does not create a magnetic field.

\subsection{Magnetic field generation in the field of a plasma wave}\label{sec33}
\begin{figure}[ht]
  \includegraphics{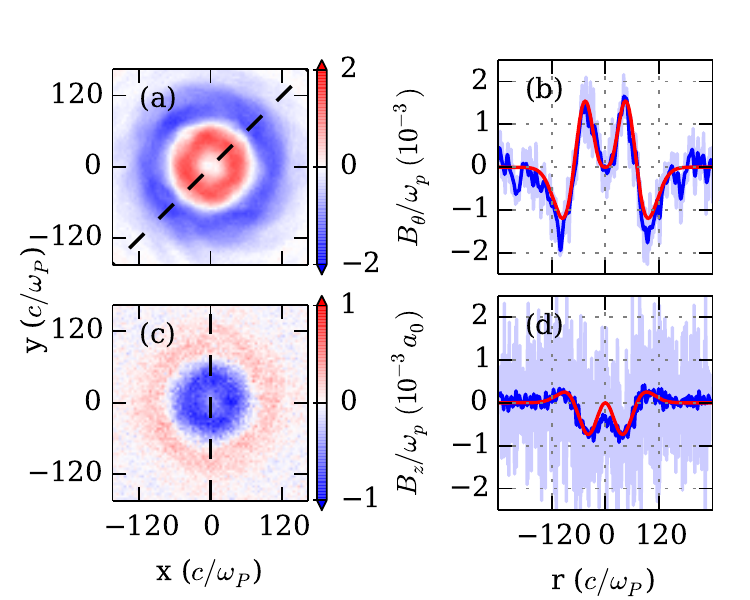}
  \caption{\label{figpicb} Results from the same PIC simulation as in Fig. \ref{figpvt}. The plots on the top show the azimuthal magnetic field $B_{\theta}$ and the bottom show the axial magnetic field $B_{z}$. The plots on the left show the central 2D transverse slice from the PIC box (filtered using a gaussian filter with $\sigma=1$ cell) and the plots on the right show line-outs from these slices (light blue unfiltered data, blue filtered data) compared with a theoretical model using $a_0 = 0.2$ (red). Due to this being a second-order effect there is considerable noise seen in the magnetic field, despite there being $\sim100$ particles per cell in the simulation. Nevertheless, after a filter is applied to the axial field, a good match to the theoretical model can be seen.}
\end{figure}

\begin{figure*}[ht]
  \includegraphics[width=0.9\textwidth]{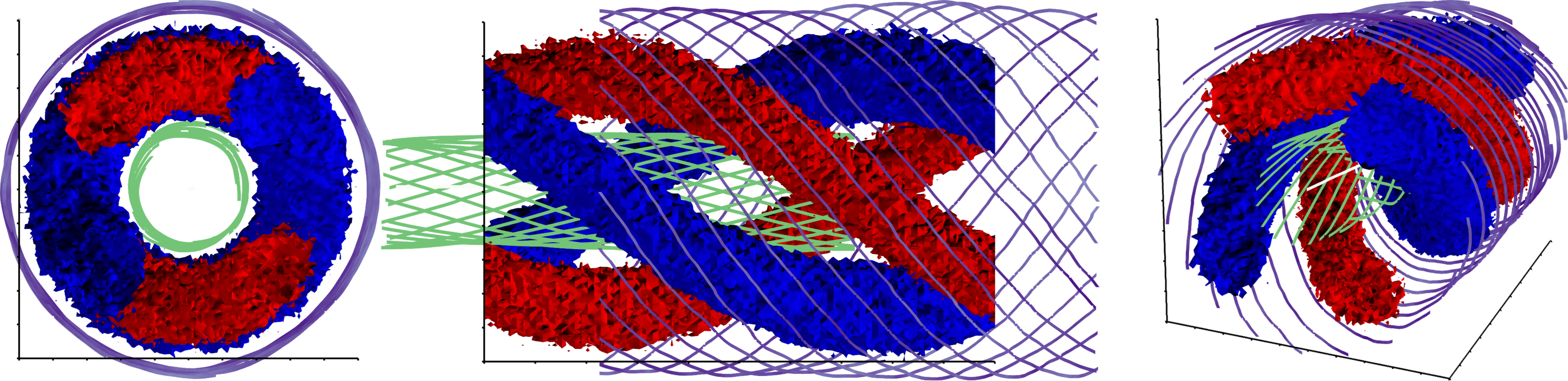}
  \caption{\label{figpicfieldlines} Three 3D views of the OCEAN PIC simulation results for the plasmon mode $p=0,\,l=2$. The view along the propagation axis, and across the transverse axis, are shown with a parallel projection, and the final tilted image uses a convergent projection. The red and blue surfaces show surfaces of constant $n_e$ at $80\%$ of the amplitude. In addition to the surfaces of constant density the magnetic field lines for the interior (green) and exterior (purple) regions of the plasmon are shown, the fields are shown slightly offset in the centre image so that the rotational component is clearly visible.}
\end{figure*}
The magnetic field generation is a second-order effect. The general expressions for the magnetic field generation by a plasma wave are derived by Bell et al. \cite{Bell88} and Gorbunov et al. \cite{Gorbunov96}. Following the approach developed in this paper, the equation for the second-order vector potential ${\bf A}^{(2)}$ can be written as
\begin{equation} \label{eq62a}
 (\partial_t^2-c^2\nabla^2+\omega_{pe}^2){\bf A}^{(2)}=\epsilon_0^{-1}{\bf j}^{(2)},
  \end{equation}
where ${\bf j}^{(2)}=-e \delta n_e {\bf v}_e=  {\bf j}\, \delta n_e/n_{e0}$ is the second-order current. According to the expressions \eqref{eq58} - \eqref{eq61}, all three components of the vector potential are generated in the second order on the plasma wave amplitude.  Explicit expressions for the vector potential can be found from Eq. \eqref{eq62a} in the paraxial approximation, accounting only for the dominant axial derivative in the Laplacian term:
\begin{eqnarray}
  &   A_z=&- \frac{ekE_{0}^2}{2m_e \omega^3}F_{p,l}^2-\frac{ekE_{0}^2}{2m_e \omega}\,\frac{F_{p,l}^2}{4k^2c^2-3\omega^2}\,\cos2(kz-\omega t +l\theta), \label{eq62d}  \\
  &&\nonumber\\
  &   A_\theta=&- \frac{e l E_{0}^2}{2m_e \omega^3w_b}X^{-1/2} F_{p,l}^2- \frac{el E_{0}^2}{2m_e \omega w}\,\frac{X^{-1/2} F_{p,l}^2}{4k^2c^2-3\omega^2}\,\cos2(kz-\omega t +l\theta),\label{eq62e}  \\
  &&\nonumber\\
  &A_r=& - \frac{eE_{0}^2}{2m_e \omega w_b}\,\frac{X^{1/2} (F_{p,l}^2)^\prime }{4k^2c^2-3\omega^2}\, \sin2(kz-\omega t +l\theta).  \label{eq62f} 
 \end{eqnarray}
It contains quasi-stationary components for the axial and azimuthal components and the component oscillating at the second harmonic. The axial component dominates and two other components are of the first order on the paraxial parameter.  The magnetic field ${\bf B}=\nabla\times{\bf A}^{(2)}$ calculation is straightforward: 
 \begin{eqnarray}
&&   B_z=- \frac{elE_{0}^2}{m_e \omega^3 w_b^2}\left(F_{p,l}^2\right)^\prime, \label{eq62}  \\
&&   B_\theta= \frac{ekE_{0}^2 }{m_e \omega^3 w_b}\,X^{1/2}(F_{p,l}^2)^\prime,\label{eq62b}  
 \end{eqnarray}
The radial component of the magnetic field is zero, $B_r=0$, and the azimuthal component dominates. It is of the first order on the paraxial parameter compared to the axial component which is of the second order. The magnetic field is constant, it does not oscillate in time and in space. Magnetic field lines form helices of a constant radius rotating in the direction opposite to the sign of the  orbital momentum: $\theta=\theta_0 -(k/l)(z-z_0)$. 

It is important to note that the total magnetic flux over any closed surface is zero: $\oint {\bf B}\cdot d{\bf S}=\int dV\,\nabla\cdot{\bf B}=0$. In particular, $\int_0^\infty B_z r\,dr=0$ and magnetic field is zero at the axis. Generation of that magnetic field is an adiabatic effect. Its intensity is proportional to the square of the plasma wave electric field and it disappears as soon as the plasma wave disappears. 

The azimuthal component of magnetic field is created by the quasi-static electric current associated with the axial momentum carried with the plasma wave. The axial component of the magnetic field can be related to the orbital momentum carried by the plasma wave. The latter can be derived from the general expression for the electromagnetic stress tensor \cite{Jackson}. In the case of a zero magnetic field it reads $T_{ij}=\epsilon_0(E_iE_j-\frac12 \delta_{ij}E^2)$. In plasma we need also to add the particle stress tensor $\sigma_{ij}=n_{e0} m_e u_i u_j$. Then the wave momentum is described by the projection of the stress tensor on the correspondent direction divided by the phase velocity, $P_j=(k/\omega)\sum_j n_i(T_{ij}+\sigma_{ij})$. The dominant term in the expression for the axial component of the orbital momentum reads:
\begin{equation} 
  L_z= r P_\theta k/\omega= r \,(k/\omega)\,(\epsilon_0 E_z E_\theta+n_{e0}m_e u_z u_\theta) =\frac{\epsilon_0 l}{\omega} E_{0}^2 F_{p,l}^2.
  \end{equation}
Correspondingly, the total orbital momentum per-unit-length carried with a twisted plasma wave is ${\mathcal L}_z=2\pi \int_0^\infty r\,dr L_z=\pi w_b^2\epsilon_0 l E_{0}^2/\omega$. (Here the relation \eqref{eq25} for the normalization of the radial functions is used.)
One then can define the magnetic moment per-unit-volume 
$$M_z=-\frac{e}{m_e} L_z=-\frac{\epsilon_0 e  l}{m_e\omega} E_{0}^2  F_{p,l}^2 $$ 
and the total magnetic moment per-unit-length ${\mathcal M}_z=-\pi w_b^2\epsilon_0 e l E_{0,1}^2/m_e\omega$. It can be noticed that one cannot obtain the same expression for the magnetic moment by applying the definition of the magnetic moment of electrons as $M_z=rj_\theta^{(2)}=rj_\theta \delta n_e/n_{e0}$. The component of magnetic moment carried with the plasma wave field is lost in such a definition.

The magnetic fields at later times in the PIC code (Fig. \ref{figpicb}) show a twisted solenoid like magnetic field which matches the theoretical calculation above. The numerical values of the magnetic field match theoretical calculations even after the plasmon is left to oscillate for $\sim20$ periods, despite being close to the noise threshold of the PIC calculation.

The plasmon described in Ref. \cite{Shi18} is a $p=0,\,l=2$ mode, which appears to have the same form as that described in this paper. The magnitude of the axial component of the magnetic field that is shown in that paper is also in agreement with the work presented here. With a plasmon of amplitude $a_0=0.2-0.3$, an OAM mode $p=0,\,l=2$,  a plasma density of $4.5\times10^{18}$ cm$^{-3}$ and a beam waist $w_{0}=5\,\mu$m the maximum amplitude in the axial direction is $2.5-5.6$ T, which is consistent with the similar conditions in Fig. 2 of Ref. \cite{Shi18}. The theoretical analysis undertaken in Ref. \cite{Shi18} involves considering ring like current structure that ignores the axial components, however this would not be sufficient to calculate the azimuthal magnetic field. No azimuthal magnetic field is shown in the numerical results, and so the full structure of the magnetic field cannot be commented on.

\section{Conclusions}\label{sec4}
This study of electrostatic electron plasma waves with orbital angular momentum covers two main aspects of the physics of these objects. The first aspect being the development of a fully kinetic paraxial perturbation applied to the electron distribution function. In contrast with Ref. \cite{Mendonca12}, the analysis presented here includes corrections due to the coupling of nearest-neighbour modes due to gradient terms in the linearized Vlasov equation. This new electron distribution function is used to develop a dispersion equation allowing for the calculation of the phase and group velocities of such a plasma wave including this coupling. The wave orbital momentum and final radial extent result in a stronger wave dispersion and stronger damping. These effects become particularly important for the long wavelength plasma waves, where $\lambda_p>kw_{0}\lambda_{De}$. It is expected that the collisionless damping of the twisted plasma wave will result in the transfer of the wave orbital momentum to the resonant electrons.

The second subject of this study is magnetic field generation by an OAM plasmon. Whilst the first-order field is shown to be zero, to second order a significant `'twisted solenoid-like'' magnetic field is shown to exist in both theoretical calculations and particle-in-cell simulations. 
\begin{acknowledgments}
This work was granted access to HPC resources of TGCC under the allocation A0010506129 made by GENCI. We acknowledge PRACE for awarding us access to resource Joliot Curie-SKL based in France at TGCC Center. The authors acknowledge support from MEPhI Academic Excellence Project (Contract No. 02.a03.21.0005-27.08.2013) and from the project ELITAS (ELI Tools for Advanced Simulation) CZ.02.1.01/0.0/0.0/16\_013/0001793 from the European Regional Development Fund.
\end{acknowledgments}

\appendix
\section{Calculation of the coupling coefficients}\label{appA}
Expression \eqref{eq34a} for the derivative of $F_{p,l}$ can be obtained by using expression \eqref{eq45} for the Laguerre polynomial. By taking a derivative of this expression one finds:  $x ( L_p^l)' =  (p+1)L_{p+1}^{l-1} - (l-x) L_p^l$.  At the same time, the Laguerre polynomial can also be expressed in a power series \cite{Grad}:  
\begin{equation} \label{eq45a}
  L_p^l(x)= \sum_{k=0}^p (-1)^k \frac{(p+l)!\, x^k}{k!\, (p-k)!\, (l+k)!} .
 \end{equation}
By taking the derivative of this expression one finds another presentation: $(L_p^l)' =- L_{p-1}^{l+1}$. Combining both expressions for the derivatives one has a relation between the Laguerre polynomials of different order:
\begin{equation} \label{eq45b}
 x L_{p-1}^{l+1}(x)=- (p+1)L_{p+1}^{l-1} + (l-x) L_p^l  .
 \end{equation}
This expression allows the presentation of the derivative of $F_{p,l}$ in the form given by Eq. \eqref{eq34a}.

The mode-coupling coefficient is given by Eq. \eqref{eq38}. The coefficients entering in this expression contain integrals of a product of two Laguerre polynomials multiplied by a power and exponential function:
\begin{equation}\label{eq42a}
I_j(p,l;p',l')= \int_0^\infty dx\,x^j {\rm e}^{-x} L_p^l(x) L_{p'}^{l'}(x). 
\end{equation}
The method of evaluation of that integral consists in two steps. First, the Laguerre polynomial with an index $p'$ is developed in a power series according to Eq. \eqref{eq45a}: 
\begin{equation}\label{eq43}
I_j(p,l;p',l')=\sum_{k=0}^{p'}   (-1)^k \frac{(p'+l')!}{k!\, (p'-k)!\, (l'+k)!}  \int_0^\infty dx\,x^{j+k} {\rm e}^{-x} L_p^l(x).
\end{equation}
The remaining integral can be calculated as follows: In the case $0\leq j+k\leq l-1$ its value is given in Ref. \cite{Grad} (Eq. 7.414.11):
$$ \int_0^\infty dx\,{\rm e}^{-x}x^{j+k} L_p^{l}(x)=\frac{(j+k)!\,(l+p-j-k-1)!}{p!\,(l-j-k-1)!}. $$ 
In the case  $j+k\geq l$ the remaining Laguerre polynomial $L_p^l$ can be represented according to Eq. \eqref{eq45}:
$$ \int_0^\infty dx\,{\rm e}^{-x}x^{j+k} L_p^{l}(x)=\frac1{p!} \int_0^\infty dx\,x^{j+k-l} d_x^p \left({\rm e}^{-x} x^{l+p}  \right). $$ 
The integral in the right hand side is calculated by integrating it by parts $p$ times.  It has non-zero value only if $j+k\geq l+p$:
$$ \int_0^\infty dx\,{\rm e}^{-x}x^{j+k} L_p^{l}(x)= (-1)^p \frac{(j+k)!\,(j+k-l)!}{p!\, (j+k-l-p)!} .$$
Inserting these expressions in Eq. \eqref{eq43} we find a representation for the integral $I_j(p,l;p',l')$ as a finite sum:
\begin{eqnarray}\label{eq43b}
&& I_j(p,l;p',l')= \nonumber \\
&& \sum_{k=0}^{p'} (-1)^{k} \frac{(p'+l')!\,(j+k)!}{p!\,k!\, (p'-k)!\, (l'+k)!}\times \left\lbrace
\begin{array}{cc}
\frac{(p+l-j-k-1)!}{(l-j-k-1)!}, &\qquad k\leq l-j-1,\\
0,& \qquad l-j\leq k \leq p+l-j-1,\\
(-1)^p\frac{(j+k-l)!}{(j+k-p-l)!},& \qquad k\geq p+l-j.\\
\end{array}. \right.
\end{eqnarray}
By using that expression for $I_j(p,l;p',l')$ one can calculate explicitly the expressions for the coefficients $K^{\pm}$. Because of the symmetry $I_j(p,l,p',l')=I_j(p',l',p,l)$, the summation limits $p+l-j\leq p'$ and $p'+l-j\leq p$ need $p'+l-j \geq p \geq p'-l+j$ and $l+l'\leq2 j$. Then the non-zero coefficients needed for calculation of the coupling coefficient $Q_{pp}$ are given by Eqs. \eqref{eq46a} and \eqref{eq46b}.

\section{Numerical modeling}\label{appB}
There are several ways that a plasma oscillation can be independently generated in a PIC simulation: by imposing a perturbation on the electron distribution function and solving the Gauss equation for a potential; or by imposing an electric field of the correct form to generate the electron distribution function associated with the desired plasmon. The latter method has the advantage that it can be performed gradually in time. Since one of the principle aims of the simulation presented here is to demonstrate the existence of certain second order effects, a gradual onset method gives the system time to relax into a stable state.

A small perturbative electric field is imposed volumetrically each time-step over ten plasma periods $T_{pe}=2\pi/\omega_{pe}$, the amplitude of this field $a_0=0.3/10T_{pe}$ is such that if the process is $100\%$ efficient then a plasmon with an $a_0=0.3$ will be generated. However there is an impedence to the system due, at least in part, to numerical noise, thermal effects, and the positioning of the absorbing boundaries, so the plasmon generated has an amplitude some fraction of the targeted amplitude. To achieve a closer match to the stated $a_0=0.3$ a larger box is required, but given that this requires additional computational time a compromise is made.

\begin{figure}[ht]
  \includegraphics{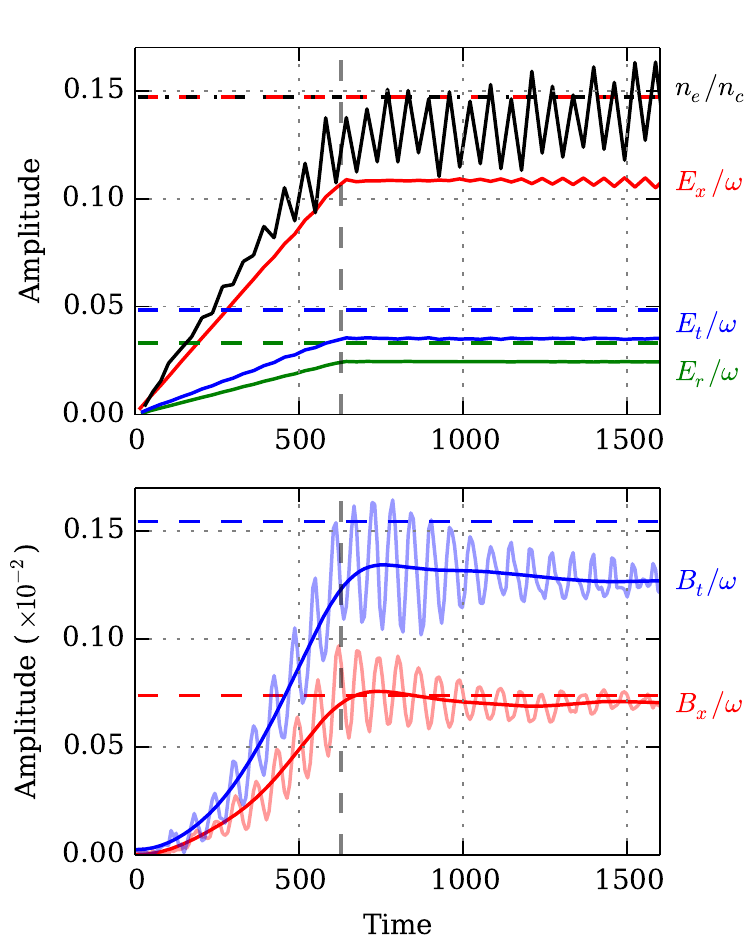}
  \caption{\label{figbtime} The upper plot shows the amplitude of the electric field components, with red showing the axial, blue the azimuthal, green the radial electric fields, and finally black showing the electron density component. The vertical dashed line shows the point in the simulation where the amplification process stops. The horizontal dashed lines show the component amplitudes for a reference $a_0=0.2$ in the same color scheme. The lower plot shows the amplitude of magnetic field components in the PIC simulation. Red lines correspond to the axial field, and blue to the azimuthal field, with the solid line corresponding to the PIC simulation, the darker solid line corresponds to a average over a Gaussian window over 3 periods. No radial magnetic field is observable.}
\end{figure}

The plasmon generated in the simulation presented here attains an amplitude in the range $0.1<a_0<0.2$. The electric field across all components achieved a consistent amplitude of $a_0=0.15\pm0.01$. The electron density component of the plasmon has some non-linear features where the positive density part of the wave has a slightly larger amplitude than the negative density part, both the negative and positive amplitudes correspond to an $a_0 $ in the range $0.2>a_0>0.15$. The amplitudes of the various fields over time, plotted normalized to $eE_c/m_ec\omega_p$, $eB_c/m_e\omega_p$ and $n_e/n_c$, can be seen in Fig. \ref{figbtime}. The frequency of the plasmon matches that described by the dispersion relation Eq. \eqref{eq43}, which in this case reduces to the simple result $\omega\simeq\omega_{pe}$, with all electric and density components oscillating with the same frequency. The single plasmon observable in the simulation is stable for a time greater than 20 oscillations after initial 10 oscillation set-up phase.

To run a simulation with a grid consisting of $1200\times1200\times160$ cells and 100 particles per cell, approximately 24 hours on 20000 cores is required. Such a high resolution is required for several reasons; the first reason being that the temperature is required to be low to avoid such effects as wave-breaking and landau damping; the second reason being that when a lower resolution is used the amplification process is less efficient as the observed plasmon becomes out of phase with the amplifying field; thirdly the second-order magnetic fields are not observable in conditions with greater noise. If the grid size is reduced by a factor of 4 and the number of particles reduced to just 10 per cell a plasma wave of the same amplitude here is still observable.

The amplitude of the two components of the magnetic field correspond to an $a_0 $ in the range $0.2>a_0>0.15$ consistent with the amplitude observed in the electron density component. The profile of the magnetic field, while static in space, has some temporal oscillation (see Fig. \ref{figbtime}) at a frequency of $\omega_{pe}$ during the amplification phase but with a small amplitude of around $10-20\%$ of the mean value and decaying in time towards an equilibrium value. At very late in time an oscillation at a frequency of $2\omega_{pe}$ is visible in the azimuthal field.

\section{Electron distribution function of the vortical mode (0,\,2)}\label{appC} 
Here we present an example of the electron distribution function for an LG plasma wave, considered in Section \ref{sec3}, of the mode $p=0$, $l=2$, and for simplification the Gouy phase and front curvature are ignored.  The corresponding radial functions read:
$$ F_{0,2} = X/\sqrt{2}\,{\rm e}^{-X/2}, \qquad F_{1,1} =\sqrt{X/2}\,(2-X)\,{\rm e}^{-X/2}.$$
The electric potential \eqref{eq29} contains only one term characterized by the amplitude  $\phi_{0,2}$:
$$  \Phi= \frac{\phi_{0,2}}{\sqrt{2}}\frac{r^2}{w^2_b} {\rm e}^{-r^2/2w^2_b} \,\cos(kz-\omega t +2\theta).     $$
The electric field can then be found by taking the gradient of the potential.
The dominant term in the expansion of the electron distribution function \eqref{eq30} is given by Eq. \eqref{eq39a}. In the first-order expansion over the paraxial parameter $1/kw_b\ll1$, the expression is straightforward:
$$  f_{0,2} = \frac{k v_z}{\omega-kv_z} e\phi_{0,2} \partial_{\varepsilon}f_{e0}. $$
Other coefficients are of the first order. They are following from Eq. \eqref{eq38a} by taking into account the fact that non-zero coefficients are given by Eqs. \eqref{eq46a} and \eqref{eq46b}. The three components of the electron distribution function in the first order are the following:
\begin{eqnarray}
&f_{0,2\pm1}&=  \frac{iv_\perp}{2w}\frac{\omega}{(\omega-kv_z)^2} e\phi_{0,2} \partial_{\varepsilon}f_{e0}\, {\rm e}^{\mp i\theta_v}\sqrt{2+\frac{1\pm1}{2}}, \nonumber \\
  &f_{1,0}&=-\frac{iv_\perp}{2w}\frac{\omega}{(\omega-kv_z)^2}  e\phi_{0,2} \partial_{\varepsilon}f_{e0}\, {\rm e}^{i\theta_v}. \nonumber
  \end{eqnarray}
With these expressions one can calculate the explicit form of the electron distribution function:
\begin{eqnarray}
  \frac{\delta f_e}{f_{e0}}=&&-a_0 \frac{kc^2}{\omega v_{th}^2}\,\frac{\omega}{\omega-kv_z}\,\frac{r}{w_b}\,{\rm e}^{-r^2/2w^2_b}\,\bigg[v_z\frac{r}{w_b}\cos(kz-\omega t+2\theta) +\nonumber \\
    &&+\frac{v_\theta}{kw_b}\frac{\omega}{\omega-kv_z}\cos(kz-\omega t+2\theta) +\frac{v_r}{kw_b}\frac{\omega}{\omega-kv_z} \bigg(\frac{r^2}{w^2_b}-2\bigg)\sin(kz-\omega t+2\theta)\bigg].\nonumber
\end{eqnarray}
Here the dimensionless amplitude $a_{0}=e\phi_{0,2}/\sqrt{2}m_e c^2$ is introduced. This expression can be used for calculation of the moments of the electron distribution function.

\end{document}